\documentclass[a4paper,12pt] {article}
\pdfoutput=1 

\usepackage{jheppub} 
\usepackage{amsthm,enumerate}
\usepackage[utf8]{inputenc}
\usepackage{lmodern}
\allowdisplaybreaks
\newcommand{\mss}{\kern 1pt}

\newcommand{\tends}[1]{\bbuildrel{\hbox to 2em{\rightarrowfill}}_{#1}^{}}

\newcommand{\Int}[1]{\,\mathop{\!#1}\limits^{\lower1ex\hbox{$\scriptstyle\circ$}}{}}

\theoremstyle{remark}

\newcommand {\bp}{\begin{pmatrix}}
\newcommand {\nn}{\nonumber}
\newcommand {\ep}{\end{pmatrix}}
\newcommand{\be}{\begin{equation}} \newcommand{\ee}{\end{equation}}
\newcommand{\bea}{\begin{eqnarray}}\newcommand{\eea}{\end{eqnarray}}
\def\clap#1{\hbox to 0pt{\hss#1\hss}}

\title{Integrable coupled bosonic massive Thirring model and its nonlocal reductions}
\author[a]{B. Basu-Mallick,}
\author[b,1]{Debdeep Sinha, \note{Corresponding author.}}
\affiliation[a]{Departamento de F\'\i sica Te\'orica, Universidad Complutense de Madrid, Plaza de las Ciencias 1, 28040 Madrid, SPAIN}
\affiliation[b]{Theory Division, Saha Institute of Nuclear Physics, HBNI, 1/AF Bidhannagar,\\
  Kolkata 700064, INDIA}

\emailAdd{bireswar.basumallick@saha.ac.in} 
\emailAdd{debdeep.sinha@saha.ac.in}

\abstract{%
  A coupled bosonic massive Thirring model (BMTM), involving an interaction between the two independent 
spinors, is introduced and shown to be integrable. By incorporating suitable reductions between the field components of the 
coupled BMTM, five novel integrable models with various type of nonlocal interactions are constructed. 
Lax pairs satisfying the zero curvature 
condition are obtained for the coupled BMTM and for each of the related nonlocal models. 
An infinite number of conserved quantities are derived for each of these models which confirms the integrability 
of the systems. It is shown that the coupled BMTM respects important symmetries of the original BMTM
 such as parity, time reversal, global $U(1)$-gauge and the proper Lorentz transformations. Similarly, all the nonlocal 
models obtained from the coupled BMTM remain invariant under combined operation of parity and time reversal transformations. However, 
 it is found that only one of the nonlocal models is invariant under proper Lorentz transformation and two other models are invariant under 
global $U(1)$-gauge transformation. By using ultralocal Poisson bracket relations among the elements of the 
Lax operator, it is shown that the coupled BMTM and one of the nonlocal models are completely integrable in the Liouville sense.
}

\begin{document}



\maketitle
\flushbottom

\section{Introduction}

The massive Thirring model (MTM) was originally introduced 
in $1+1$ dimension in the context of quantum field theory as a soluble toy model to understand various 
aspects of non perturbative physical phenomena relevant in the realistic $3+1$ dimension \cite{th}. 
At the classical level, it is possible to consider two versions of this MTM. The first version is obtained by considering the fields 
as functions taking value in the complex number field and the second one is obtained by considering the fields as functions taking value 
in the Grassmann algebra \cite{kuli,kuli1}. The first case often refers to as the bosonic massive Thirring model (BMTM), which has some 
important physical applications. For example, the BMTM describes the nonlinear optical pulse propagation 
in Bragg nonlinear optical media \cite{e1,e2,e3,e4,e5}. The BMTM also appears in case of nonlinear atomic 
optics in order to describe the dipole-dipole interaction among many-body bosonic atoms \cite{prl71}. 
Further importance of this model arises due to the remarkable fact that this model is completely integrable. 
The classical integrability of this model was investigated by Mikhailov \cite{avm} and also independently by 
Orfanidis \cite{r1}. This model has been studied extensively by employing inverse scattering transformation 
method, Backlund  and Darboux  transformations and various types of soliton solutions are obtained 
\cite{avm,r1,r2,r3,r4,wa, ps, r5, r6}. The rogue wave solution for this model has been obtained \cite{rw2,rw3,rw4} and  
soliton solutions for BMTM in a nonvanishing background has also been considered  \cite{ig1,ig2,ig3}. 
The connection between this model with other integrable models such as sine Gordon equation and derivative 
nonlinear Schr$\ddot{\rm o}$dinger equation has been discussed in the literature \cite{r1,bm}. Recently, general bright and dark soliton solutions 
to BMTM has been obtained via KP hierarchy reductions \cite{cf}. Another interesting development in this 
context is the study of integrability of MTM in the presence of defect through inverse scattering transformation 
method \cite{agu, agu1,agu2}.  Apart from this, some generalized version of BMTM which includes balanced loss and 
gain has been considered \cite{ig4}. Quantum integrability and exact solvability of fermionic and bosinic versions of MTM
have also been studied by using coordinate and algebraic Bethe ansatz \cite{thac, kul, ta}.

In the present investigation, we introduce a coupled BMTM as a classical field theoretic model in which case two spinors interact with
each other. The Lagrangian and Hamiltonian formulation for this coupled BMTM are
presented. This system is Lorentz invariant and respects other important symmetries, such as parity, time reversal
and global $U(1)$ gauge symmetry of the original BMTM. 
The Lax pair for the coupled BMTM is constructed
and the equation of motions are obtained as a zero curvature condition. By using the zero curvature condition it is shown that the system possesses
an infinite number of local conserved quantities, which confirms the integrability of the system. Furthermore, in order to 
prove the complete integrability of the system,  the ultralocal Poisson bracket (PB) relations among the elements of the Lax operator
is calculated and it is shown that the corresponding monodromy matrix satisfies the classical Yang-Baxter equation. It follows from 
this construction that the conserved quantities of coupled BMTM
are in involution and hence the system is completely integrable in the Liouville sense.

An important outcome of the present investigation is that  various consistent reductions between the field components of the 
coupled BMTM can be used to reproduce the original BMTM and to generate new integrable systems. Indeed, the original BMTM can be obtained
from the Lax pair and zero curvature formulation of coupled BMTM under a specific reduction between the field components.
Moreover, in the present paper we find out a new set of nonlocal reductions between the field components
of the coupled BMTM, that generate five types of novel nonlinear integrable field models in $1+1$ dimension characterized by various 
types of nonlocal interactions. The resulting interactions are nonlocal in the following three senses;
i) Space inversion: In this case the value of the interacting potential at $(x,t)$ requires the information on the fields at $(x,t)$ 
as well as at $(-x,t)$, ii) Time inversion: In which case the value of the interacting potential at $(x,t)$ requires the information on the fields
at $(x,t)$ as well as at $(x,-t)$ and finally, iii) Space-time inversion:  In this case the value of the interacting potential at $(x,t)$ requires the information 
on the fields both at $(x,t)$ and $(-x,-t)$.

It should be mentioned here that a nonlocal integrable model
with space inversion
has been originally studied in Ref \cite{ab} in the context of nonlinear Schr$\ddot{\rm o}$dinger equation (NLSE). 
Furthermore, it is shown that the nonlocal NLSE possesses an infinite number of conserved quantities confirming the integrability of the system and the
inverse scattering transformation method is employed to obtain soliton solutions. The drak and bright soliton 
solitons for such system are also obtained \cite{n0}. The nonlocal NLSE and its various generalizations 
are considered in the literature \cite{n2, n1,n1a, n3,n5,1ds}. Subsequent developments  in the study of nonlocal integrable models 
lead to new classes of nonlocal derivative NLSE \cite{dnl,dnl1} and nonlocal sine-Gordon model as well \cite{sg1}. 


The nonlocal systems considered in the present investigation may be treated as nonlocal variants of  BMTM in the sense that 
the original BMTM and the five nonlocal models  can be obtained from the same generic Lax pair of the coupled BMTM 
containing four independent fields. 
It turns out that all the nonlocal models are symmetric under combined operation of parity ${\cal P}$ and time reversal ${\cal T}$
transformations while three of them remain invariant separately under parity ${\cal P}$ and time reversal ${\cal T}$.
Furthermore, one of the models remains invariant under 
proper Lorentz transformation and two other models respect the global $U(1)$ gauge symmetry. By using the equations for the Jost functions 
and their corresponding compatibility conditions, an infinite number of conserved quantities are obtained for each of the nonlocal models
which confirms the integrability of the nonlocal systems. The Hamiltonian structure and consistent
PB relations are obtained for the space inverted  nonlocal systems. The 
complete integrability of one of the nonlocal models with space inversion is discussed and it is shown that various elements of the monodromy matrix 
for this system satisfy the same PB relations among themselves as in the case of the coupled BMTM.
From this result it follows that the conserved quantities associated with this system are in involution, which confirms the complete integrability 
of this space inverted nonlocal model in the Liouville sense.

The plan of this paper is as follows. The coupled BMTM, involving an interaction between two spinors, is introduced in the next section.
The Lax pair for this model is constructed and the equations of motions are obtained as a zero curvature condition. Five new nonlocal
integrable models, obtained via various nonlocal reductions among the field components of the coupled BMTM, are constructed in subsection 2.1. Subsequently, various symmetry properties 
 of the coupled BMTM and all the nonlocal systems are presented in the subsection 2.2. 
The conserved 
quantities associated with the coupled BMTM and the related nonlocal models are derived by using the zero curvature formulation in section 3. 
The Hamiltonian formulation for the coupled 
BMTM and two of the related models with space inverted nonlocality are presented in section 4.  Physical significance of few
conserved quantities associated with coupled BMTM and related nonlocal systems are also discussed in this section. The complete integrability of the coupled BMTM and one of the nonlocal models are established in section 5. 
In the last section, we summarize the results obtained in this paper and discuss some of future prospective.

\section{Introduction to the models and the construction of the Lax matrices}

The Thirring model is a well known relativistic exactly solvable nonlinear model in $1+1$ dimension and has the following form,
\bea
i\gamma^{\mu}\partial_{\mu}\psi=m\psi-g J_{\mu}\gamma^{\mu}\psi,\ \ \ J^{\mu}=\bar{\psi}\gamma^{\mu}\psi
\label{1tm}
\eea
where $\psi=(\psi_1, \psi_2)^T$ with $T$ denotes a transpose, $J_0=J^0$, $J_1=-J^1$, $\bar{\psi}=\psi^{\dagger}\gamma^0$, 
with `$\dagger$' denotes transpose plus complex conjugation,
$\gamma^0=\sigma_1$, $\gamma^1=-i\sigma_2$ and $\sigma_i, i=1,2,3$ are Pauli matrices, $m$ denotes the mass and $g$ 
 is the coupling constant. Earlier 
studies on the integrable properties of  this model may be found in Refs. \cite{avm, r1}. For $g=0$, Eq. (\ref{1tm}) represents 
the free Dirac equation in $1+1$ dimension. However for $g\ne0$, the last term in Eq. (\ref{1tm}) incorporates a self interaction 
of the field $\psi$ and introduces a nonlinear coupling between the field components. 

In the present investigation, we introduce a coupled BMTM in  $1+1$ dimension that incorporates the interaction between two Lorentz spinours $\psi$ and $\phi$. The Lagrangian density for such coupled BMTM is expressed in the following manner,
\bea
{\cal L}&=&\frac{1}{2}\bar{\phi}\big(i\gamma^{\mu}\partial_{\mu}-m\big)\psi+\frac{1}{2}\bar{\psi}\big(i\gamma^{\mu}\partial_{\mu}-m\big)\phi-\frac{1}{2}\bar{\phi}\big(i\gamma^{\mu}\overset {\leftarrow}\partial_{\mu}+m\big)\psi 
-\frac{1}{2}\bar{\psi}\big(i\gamma^{\mu}\overset {\leftarrow}\partial_{\mu}+m\big)\phi \nn \\
&+&\frac{g}{2}\big[(\bar{\phi}\gamma_{\mu}\psi)(\bar{\phi}\gamma^{\mu}\psi)+(\bar{\psi}\gamma_{\mu}\phi)(\bar{\psi}\gamma^{\mu}\phi)\big],
\label{L}
\eea 
where $\phi=(\phi_1, \phi_2)^T$,  $\bar{\phi}=\phi^{\dagger}\gamma^0$. 
The last term within the square bracket represents the interaction between the fields $\psi$ and $\phi$ 
where the strength of the coupling is determined by the real coupling constant $g$. 
The equations of motion for the coupled BMTM, can easily 
be derived form the Lagrangian density in Eq. (\ref{L}) using Euler-Lagrangian equation. The variations of the  Lagrangian density in Eq. (\ref{L})
with respect to $\bar{\phi}$ and $\bar{\psi}$ respectively produce the equations for the fields $\psi$ and $\phi$ with
the following form
\bea
&&\big(i\gamma^{\mu}\partial_{\mu}-m\big)\psi+g (\bar{\phi}\gamma_{\mu}\psi)\gamma^{\mu}\psi=0, 
\label{eq1}\\
&&\big(i\gamma^{\mu}\partial_{\mu}-m\big)\phi+g (\bar{\psi}\gamma_{\mu}\phi)\gamma^{\mu}\phi=0.
\label{eq2}
\eea
The conjugate equations, obtained by varying the Lagrangian density in Eq. (\ref{L}) with respect to $\phi$ and $\psi$,
are respectively given by,
\bea
&&\bar{\psi}\big(i\gamma^{\mu}\overset {\leftarrow}\partial_{\mu}+m\big)-g{\bar\psi} \gamma^{\mu}(\bar{\psi}\gamma_{\mu}\phi)=0,
\label{eq3}\\
&&{\bar\phi}\big(i\gamma^{\mu}\overset {\leftarrow}\partial_{\mu}+m\big)-g{\bar\phi} \gamma^{\mu}(\bar{\phi}\gamma_{\mu}\psi)=0.
\label{eq4}
\eea
It should be noted that for the special case $g=0$, each of the fields $\psi$ and $\phi$ satisfy the free Dirac equation. However, in this special case the  Lagrangian  density (\ref{L}) does not reduce to the usual Lagrangian density of two free fermions. This is because the kinetic and the mass terms in this Lagrangian density involve a coupling between two spinors.
Before proceeding further, here we employ the following scale transformations
\bea
x\rightarrow \frac{1}{m}x,\  t\rightarrow \frac{1}{m}t,\ \psi_j\rightarrow \sqrt{\frac{2m}{|g|}}\psi_j, \  \psi_2, \ \phi_j\rightarrow \sqrt{\frac{2m}{|g|}}\phi_j, \ j=1,2; \ \epsilon=-\text{sgn}(g),
~\label{cv}
\eea
and write Eqs. (\ref{eq1}) and (\ref{eq2}) in the component form which respectively takes the form,
\bea
&& i\big(\partial_t+\partial_x\big)\psi_1-\psi_2-4\epsilon (\psi_2\phi_2^{*})\psi_1=0,\nn\\
&& i\big(\partial_t-\partial_x\big)\psi_2-\psi_1-4\epsilon (\psi_1\phi_1^{*})\psi_2=0,
\label{qs}
\eea
and 
\bea
&& i\big(\partial_t+\partial_x\big)\phi_1-\phi_2-4\epsilon (\phi_2\psi_2^{*})\phi_1=0,\nn\\
&& i\big(\partial_t-\partial_x\big)\phi_2-\phi_1-4\epsilon (\phi_1\psi_1^{*})\phi_2=0.
\label{rs}
\eea
After the scale transformation, the conjugate equations (\ref{eq3}) and (\ref{eq4}) in the component form can easily be obtained 
by taking the complex conjugation of (\ref{qs}) and (\ref{rs}). It is convenient for later purpose
to write the original BMTM in Eq. (\ref{1tm}) in the component form as,
\bea
&&i \left(\partial_t+\partial_x\right)\psi_1-\psi_2-4 \epsilon(\psi_2\psi^*_2) \psi_1=0\nn\\
&&i \left(\partial_t-\partial_x\right)\psi_2-\psi_1-4 \epsilon (\psi_1\psi^*_1)\psi_2=0
\label{tmc}
\eea
where the scaling (\ref{cv}) is employed. 

In order to study the integrable properties of the coupled BMTM, here we  employ the 
zero curvature formulation. In this formulation, it is crucial to construct the Lax pair, i.e $U$, $V$ matrices for the system. 
The Lax pair, in general, is defined in terms of a pair of linear equations of the following form:
\bea
v_x=Uv; \ \ v_t=Vv,
\label{uvn}
\eea
where $v_x=\frac{dv}{dx}$, $v_t=\frac{dv}{dt}$ and $v$ is column vector. For example, in the present case
 $U$, $V$ are $2\times 2$ matrices and  $v$ is of the form 
$v=(v_1, v_2)^{T}$. The compatibility condition $v_{xt}=v_{tx}$ 
arising from Eq. (\ref{uvn}) yields the following matrix equation:
\bea
U_t-V_x+[U,V]=0,
\label{zc}
\eea
which is called the zero curvature condition.
In the present investigation we take $U$, $V$ matrices for the coupled BMTM, represented by Eqs. (\ref{qs}) and (\ref{rs} in the
following manner
\bea
U&=&
\bp
i\epsilon \rho_--\frac{i}{4}(\lambda^2-\frac{1}{\lambda^2}) & i\epsilon(\lambda \phi^*_2-\frac{\phi^*_1}{\lambda})\\
i(\lambda \psi_2-\frac{\psi_1}{\lambda}) & -i\epsilon\rho_{-}+\frac{i}{4}(\lambda^2-\frac{1}{\lambda^2})
\ep,
\label{u}\\
\nn \\ \nn \\
V&=&
\bp
i\epsilon\rho_{+}-\frac{i}{4}(\lambda^2+\frac{1}{\lambda^2}) & i\epsilon(\lambda \phi^*_2+\frac{\phi^*_1}{\lambda})\\
i(\lambda \psi_2+\frac{\psi_1}{\lambda}) & -i\epsilon \rho_{+}+\frac{i}{4}(\lambda^2+\frac{1}{\lambda^2})
\ep
\label{v}
\eea
where $\rho_{\pm}=(\phi^{*}_2\psi_2\pm \phi^{*}_1\psi_1)$ and $\lambda$ is a spectral parameter independent of $x$ and $t$.
For this Lax pair, Eq. (\ref{zc}) in the component form produces the coupled BMTM as given by Eqs.(\ref{qs}) and (\ref{rs}). Therefore, 
we obtain the Lax pair associated with two sets of linear equations as given by Eqs.(\ref{uvn}) and 
the coupled BMTM is obtained via the zero curvature condition given by Eq.(\ref{zc}).


\subsection{Nonlocal integrable models related to BMTM}

 In this section, our aim is to construct new nonlocal integrable models related to BMTM by taking various reductions
between the field components.
 It should be noted that so far 
$\psi$ and $\phi$ are treated as independent spinors each satisfying the free Dirac equation in the absence of
 any interaction. The coupled BMTM, given by Eqs. (\ref{qs}) and (\ref{rs}), arises when these two fields interact in a certain 
fashion. However, it is possible to introduce reductions of definite type between the components of spinors $\psi$ and $\phi$ 
that generate new integrable models. To this end, at first we may note that the original BMTM in Eq. (\ref{tmc}) can be reproduced
 from Eqs. (\ref{qs}) and (\ref{rs}) under the reduction:
\bea
\bp
\phi_1\\
\phi_2
\ep
= \bp 1 & 0\\
0 & 1
\ep
\bp
\psi_1(x,t)\\
\psi_2(x,t)
\ep.
\label{r0}
\eea
Moreover, the Lax pair for the original BMTM can be obtained by substituting the relation (\ref{r0}) in 
Eqs.\,(\ref{u}) and (\ref{v}). In a similar way, under various 
nonlocal reductions between the fields $\{\psi_1, \psi_2\}$ and $\{\phi_1, \phi_2\}$, Eqs. (\ref{qs}) and (\ref{rs}) give rise to the new
nonlinear integrable equations and Eq. (\ref{u}) and (\ref{v}) yields the corresponding Lax pair. A systematic investigation of the reductions between the fields $\{\psi_1, \psi_2\}$ and 
$\{\phi_1, \phi_2\}$ that give rise to the new sets of nonlinear systems, is one the main concern of the present 
investigation. A list of these new reductions along with the corresponding nonlocal nonlinear integrable models is presented below.\\

\noindent a) Type-I: Real reverse space: In this case the reduction is given by the relation
\bea
\bp
\phi_1\\
\phi_2
\ep
= \bp 0 & 1\\
1& 0
\ep
\bp
\psi_1(-x,t)\\
\psi_2(-x,t)
\ep.
\label{r1}
\eea
The corresponding nonlinear equation, which follows from Eqs. (\ref{qs}) and (\ref{rs}) under 
the reduction (\ref{r1}), has the form,
\bea
&&i \left(\psi_{1t}+\psi_{1x}\right)- \psi_2-4 \epsilon \psi_1(\psi_2\psi^*_1(-x,t))=0,\nn\\
&&i \left(\psi_{2t}-\psi_{2x}\right)-\psi_1-4 \epsilon \psi_2(\psi_1\psi^*_2(-x,t))=0
\label{qse1}
\eea
In the above equations and afterwards, the arguments of the fields should be taken as $(x,t)$ if it is not shown explicitly.
The Lax pair for this nonlocal model can be obtained by substituting the relation (\ref{r1}) in Eqs. (\ref{u}) and (\ref{v}).\\

\noindent b) Type-II: Real reverse time: The reduction for this type is given by the relation:
\bea
\bp
\phi_1\\
\phi_2
\ep
= \bp 0 & -1\\
1& 0
\ep
\bp
\psi_1(x,-t)\\
\psi_2(x,-t)
\ep.
\label{r2}
\eea
The Lax pair for this nonlocal system can be obtained by substituting the relation (\ref{r2}) in Eqs. (\ref{u}) and (\ref{v}). The corresponding nonlocal nonlinear equation is obtained from Eqs. (\ref{qs}) and (\ref{rs}) under the reduction (\ref{r2}) and is given by,
\bea
i \left(\psi_{1t}+\psi_{1x}\right)- \psi_2-4 \epsilon \psi_1(\psi_2\psi^*_1(x,-t))=0,\nn\\
i \left(\psi_{2t}-\psi_{2x}\right)-\psi_1+4 \epsilon \psi_2(\psi_1\psi^*_2(x,-t))=0.
\label{qse2}
\eea

\noindent c) Type-III: Complex reverse time: This type of equation is realized by the reduction:
\bea
\bp
\phi_1\\
\phi_2
\ep
= \bp 0 & 1\\
1& 0
\ep
\bp
\psi^*_1(x,-t)\\
\psi^*_2(x,-t)
\ep.
\label{r3}
\eea
and the corresponding nonlocal nonlinear equations has the following form,
\bea
i (\psi_{1t}+\psi_{1x})-\psi_2-4 \epsilon \psi_1(\psi_2\psi_1(x,-t))=0,\nn\\
i (\psi_{2t}-\psi_{2x})- \psi_1-4\epsilon \psi_2(\psi_1\psi_2(x,-t))=0.
\label{qse3}
\eea
 As in the earlier cases, the Lax pair for this nonlocal model can be obtained by substituting the relation (\ref{r3}) in Eqs. (\ref{u}) and (\ref{v}).\\

\noindent d) Type-IV: Complex reverse space: The reduction in this case is given by:

\bea
\bp
\phi_1\\
\phi_2
\ep
= \bp 0 & -1\\
1 & 0
\ep
\bp
\psi^*_1(-x,t)\\
\psi^*_2(-x,t)
\ep,
\label{r4}
\eea
and the corresponding nonlinear equation follows from Eqs. (\ref{qs}) and (\ref{rs}) and has the following expression,
\bea
i \left(\psi_{1t}+\psi_{1x}\right)-\psi_2-4 \epsilon \psi_1(\psi_2\psi_1(-x,t))=0,\nn\\
i \left(\psi_{2t}-\psi_{2x}\right)-\psi_1+4 \epsilon \psi_2(\psi_1\psi_2(-x,t))=0.
\label{qse4}
\eea
The Lax pair for this nonlocal model can be obtained by substituting the relation (\ref{r4}) in Eqs. (\ref{u}) and (\ref{v}).\\

\noindent e) Type-V: Complex reverse space-time: In this case the reduction reads:
\bea
\bp
\phi_1\\
\phi_2
\ep
= \bp 1 & 0\\
0 & 1
\ep
\bp
\psi^*_1(-x,-t)\\
\psi^*_2(-x,-t)
\ep.
\label{r5}
\eea
The corresponding nonlocal nonlinear equation is obtained from Eqs. (\ref{qs}) and (\ref{rs}) and has the following form,
\bea
i \left(\psi_{1t}+\psi_{1x}\right)-\psi_2-4 \epsilon \psi_1(\psi_2\psi_2(-x,-t))=0,\nn\\
i \left(\psi_{2t}-\psi_{2x}\right)-\psi_1-4 \epsilon \psi_2(\psi_1\psi_1(-x,-t))=0.
\label{qse5}
\eea
Similarly as above, Lax pair for this nonlocal model can be obtained by substituting the relation (\ref{r5}) in Eqs. (\ref{u}) and (\ref{v}).

The nonlocal models as obtained above may be considered as variants of the original BMTM in the sense 
that all the nonlocal models together with the original BMTM can be obtained from the Lax pair of coupled BMTM,
within the scheme of the zero curvature formulation, in a similar 
manner under various reductions between the pair of fields 
$\{\psi_1, \psi_2\}$ and $\{\phi_1, \phi_2\}$. It should be noted that,  the nonlocal NLSE introduced in Ref. \cite{ab} can be obtained from its local counterpart by reversing  
the spatial argument of one of the fields in the interaction term. However, in the 
present case the nonlocal systems can not be obtained from the original BMTM by simply reversing the arguments
of some of the fields in the interaction term. Indeed the nonlocal 
systems presented in this paper
involve different type of nonlinear interactions compared to the original BMTM and 
 may therefore be regarded as new integrable systems. However, they are closely 
related to the BMTM through the Lax-pair and zero curvature formulations.

\subsection{Symmetry properties of coupled BMTM and its nonlocal reductions}
It is well known that  the original BMTM in Eq. (\ref{tmc}) remains invariant under parity $(\cal P)$, 
time reversal $(\cal T)$, global $U(1)$ gauge and proper Lorentz transformations. 
For instance, it is easy to check that Eq.~(\ref{tmc}) remains invariant under parity $(\cal P)$ and time 
reversal $(\cal T)$ transformations characterized by
\bea
&&{\cal P}: x\rightarrow -x, \ t\rightarrow t ;  \ \ \ \  \psi_1 (x,t) \rightarrow \psi_2(-x,t), \ \ \ \  \psi_2(x,t) \rightarrow \psi_1(-x,t),\label{1P}\\
&&{\cal T}: t\rightarrow -t, \ x\rightarrow x ; \ \ \ \  \psi_1(x,t) \rightarrow \psi^*_2(x,-t), \ \ \ \  \psi_2(x,t) \rightarrow \psi^*_1(x,-t).
\label{pt}
\eea
The global $U(1)$ gauge transformation under which Eq. (\ref{1tm}) remains invariant may be expressed as
\bea
\psi_1\rightarrow e^{-i\theta}\psi_1,\ \  \psi_2\rightarrow e^{-i\theta}\psi_2.
\label{U}
\eea
The invariance under proper Lorentz transformation can be checked by introducing the 
light-cone coordinates  $\eta$ and $\xi$, given by
\bea
\eta=\frac{x+t}{2}, \ \  \xi=\frac{t-x}{2}.
\label{LC}
\eea
In this light-cone coordinate, the proper Lorentz transformation has the form 
\bea
\eta \rightarrow e^{\beta}\eta, \ \ \ \  \xi \rightarrow e^{-\beta}\xi,
\label{lt1}
\eea
and the fields transform as
\bea
\psi_1\rightarrow e^{\frac{\beta}{2}}\psi_1,\ \ \ \ \psi_2\rightarrow e^{-\frac{\beta}{2}}\psi_2.
\label{lt2}
\eea
Using Eq. (\ref{LC}), the BMTM in Eq. (\ref{tmc}) can be written as
\bea
&&i \psi_{1\eta}-\psi_2-4 \epsilon \psi_1(\psi_2\psi^*_2)=0,\nn\\
&&i \psi_{2\xi}-\psi_1-4 \epsilon \psi_2(\psi_1\psi^*_1)=0.
\label{tmc1}
\eea
It can easily be checked that Eq. (\ref{tmc1}) remains invariant under the proper Lorentz transformation 
characterized by Eqs. (\ref{lt1}) and (\ref{lt2}).

It is interesting to note that the coupled BMTM respects all the above mentioned symmetries of the BMTM.
For instance, it is easy to check that the coupled BMTM in Eqs. (\ref{qs}) and (\ref{rs}) remains invariant under parity,
time reversal and global $U(1)$ gauge transformations given respectively by Eqs. (\ref{1P}), (\ref{pt}), and (\ref{U}), provided the
field components $\{\phi_1,\phi_2\}$ transform in a same manner as $\{\psi_1,\psi_2\}$, 
under all these transformations. Moreover, in the light-cone coordinate, the coupled BMTM in Eqs. (\ref{qs}) 
and (\ref{rs}) respectively takes the following form:
\bea
&&i \psi_{1\eta}-\psi_2-4 \epsilon \psi_1(\psi_2\phi^*_2)=0,\nn\\
&&i \psi_{2\xi}-\psi_1-4 \epsilon \psi_2(\psi_1\phi^*_1)=0,
\label{ctmc1}
\eea
and
\bea
&&i \phi_{1\eta}-\phi_2-4 \epsilon \phi_1(\phi_2\psi^*_2)=0,\nn\\
&&i \phi_{2\xi}-\phi_1-4 \epsilon \phi_2(\phi_1\psi^*_1)=0
\label{ctmc2}
\eea
and remains invariant under proper Lorentz transformation as given by Eq. (\ref{lt1}) provided that the field 
components $\{\phi_1,\phi_2\}$ transform in a same manner as $\{\psi_1,\psi_2\}$ as given by Eq. 
(\ref{lt2}). Therefore, it turns out that the coupled BMTM respects all important symmetries of the original BMTM.

The nonlocal models considered in the present paper share some of the important symmetries, as mentioned above, of the original BMTM.
By employing Eqs.(\ref{1P}) and (\ref{pt}), we find that all the nonlocal models from type-I to type-V are invariant under combined operation of parity $(\cal P)$ 
and time reversal $(\cal T)$, and systems of type I, III and V remain invariant separately under 
parity $(\cal P)$ and time reversal $(\cal T)$ transformations. Further, systems of type-I and type-II remain invariant under 
global $U(1)$ gauge transformation as given by Eq. (\ref{U}). We also find that the only nonlocal model that remains invariant under proper 
Lorentz transformation is the type-V model. This can be checked by expressing the nonlocal model of type-V (\ref{qse5}) in the light cone coordinates as 
\bea
i \psi_{2\xi}-\psi_1-4 \epsilon \psi_2[\psi_1\psi_1(-\xi,-\eta)]=0,\nonumber\\
i \psi_{1\eta}-\psi_2-4 \epsilon \psi_1[\psi_2\psi_2(-\xi,-\eta)]=0,
\label{qqse5}
\eea
and subsequently using Eqs. (\ref{lt1}) and (\ref{lt2}). In the above equation the argument of the fields should be assumed to be $(\xi, \eta)$
 where it is not shown explicitly.
 By following similar steps as above, one can show that the systems I-IV are not invariant under 
proper Lorentz transformation. The above mentioned symmetries of the nonlocal models are listed in Table-1.

\begin{table}[h!]
\begin{center}
\begin{tabular} { | p {2 cm} || p {2 cm}|| p {2 cm}|| p {2 cm} || p { 2cm} || p { 2cm} | }
    \hline
    \multicolumn{6} { | c | }{Symmetries of the nonlocal models}\\
    \hline
    Model & Parity $ ({\cal P})$  & Time reversal $ ({\cal T})$ & $ {\cal PT} $   & global $U(1)$-gauge & Proper Lorentz \\
    \hline
    Type-I: & yes & yes & yes & yes & no \\
    Type-II & no & no & yes & yes & no \\
    Type-III & yes & yes &  yes & no & no \\
    Type-IV & no & no &  yes &no & no \\
    Type-V & yes & yes &  yes & no & yes\\
    \hline
    \end{tabular}
\caption{Symmetries of the nonlocal models compared to the original BMTM.}
\end{center}
\label{table}
\end{table}

We would like to make a comment in the following. After doing calculations of this paper we have noticed that, by using suibable  Lax pair which satisfy the zero curvature condition,
some multi-component extensions of BMTM
have been constructed in Ref. \cite{wada1}.
Such extensions of  BMTM equations
 take the following form (with $m=1$)
\bea
&&C_{1t}-C_{1x}+iD_{1}+\frac{i}{2}C_1D_2D_1=0,\nn\\
&&C_{2t}-C_{2x}-iD_{2}-\frac{i}{2}D_2D_1C_2=0,\nn\\
&&D_{1t}+D_{1x}+iC_{1}+\frac{i}{2}C_1C_2D_1=0,\nn\\
&&D_{2t}+D_{2x}-iC_{2}-\frac{i}{2}D_2C_1C_2=0,
\label{wada2d}
\eea
where $C_1$, $D_1$ are $p\times q $ matrices and $C_2$, $D_2$ are $q\times p $ matrices whose elements are some field variables.
The equations of BMTM
as given by Eqs. (\ref{tmc}) arises from Eqs. (\ref{wada2d}) for the case $p=q=1$ and with the reduction
\bea
C_1= \epsilon |g| C_2^*=2\sqrt{2}\,|g|^{\frac{1}{2}}\psi_2, \ \ \ \ 
D_1=\epsilon |g| D_2^*=2\sqrt{2}\,|g|^{\frac{1}{2}}\psi_1. 
\eea
However, apart from showing the above mentioned connection
with the original BMTM,
Eqs.(\ref{wada2d}) have not been analyzed further in Ref. \cite{wada1} for the case $p=q=1$. In particular, it has not been investigated whether Eqs.(\ref{wada2d}) can be written in a Lorentz covariant form  (and the corresponding Lagrangian density can be expressed as a Lorentz scalar) for the case $p=q=1$. In fact, one can easily check that 
Eqs.(\ref{wada2d}) can not be written in a  Lorentz covariant form 
if $\{C_1, C_2\}$ and $\{D_1, D_2\}$ are taken as components of 
two Lorentz spinors. 
However, we observe that, 
their exists a nontrivial mapping between apparently noncovariant  Eqs.(\ref{wada2d}) with $p=q=1$
and the presently considered Lorentz covariant coupled BMTM Eqs.(\ref{qs}) and (\ref{rs}) as 
\bea
&&C_1=2\sqrt{2}|g|^{\frac{1}{2}}\psi_2, \ \ \ \ C_2=2\sqrt{2}\epsilon |g|^{-\frac{1}{2}}\phi^*_2, \nn\\
&&D_1=2\sqrt{2}|g|^{\frac{1}{2}}\psi_1, \ \ \ \ D_2=2\sqrt{2}\epsilon |g|^{-\frac{1}{2}}\phi_1^*.
\eea

As it is well known, one of the main signature of the integrability of a given system is the existence of the conserved quantities. In the next section we shall exploit the Lax matrices and zero curvature formulation to obtain the conserved quantities for the coupled BMTM and related nonlocal models. This approach has twofold advantages. In addition to the case of vanishing boundary condition, 
this method generates an infinite number of conserved quantities for the periodic boundary condition as well. Further, this method does not relay on the underlying PB structure of the given system and therefore, is particularly helpful to construct the conserved quantities for the nonlocal systems for which the PB structure may not exit.

\section{Derivation of conserved quantities from zero curvature formulation}

In this section, at first we shall derive 
the conserved quantities for the coupled BMTM and then we use various reductions to produce the conserved 
quantities of the related nonlocal models. To this end we define the auxiliary functions $\Gamma_{ij}=v_i v_j^{-1}$, with 
$i\ne j$ and $i,j \in \{1,2\}$. Using Eq. (\ref{uvn}), it is easy to check that the auxiliary functions $\Gamma_{ij}$ satisfy the following Riccati equations, one for derivative with respect to $x$:
\bea
\partial_x\Gamma_{ij}=\left(U_{ij}-U_{jj}\Gamma_{ij}\right)+\sum_{k\ne j}\left(U_{ik}-\Gamma_{ij}U_{jk}\right)\Gamma_{kj},
\label{rcx}
\eea
and other for the derivative with respect to $t$:
\bea
\partial_t\Gamma_{ij}=\left(V_{ij}-V_{jj}\Gamma_{ij}\right)+\sum_{k\ne j}\left(V_{ik}-\Gamma_{ij}V_{jk}\right)\Gamma_{kj}.
\label{rct}
\eea
By exploiting Eqs. (\ref{rcx}), (\ref{rct}) and the zero curvature condition (\ref{zc}), the $j$-th conservation equation is obtained in the following form \cite{agu, abb}
\bea
\partial_t \left[U_{jj}+\sum_{j\ne i} U_{ji}\Gamma_{ij}\right]= \partial_x \left[V_{jj}+\sum_{j\ne i} V_{ji}\Gamma_{ij}\right].
\label{CE}
\eea
From the above equation, it follows that for vanishing boundary condition, i.e., in case the fields vanish rapidly as $|x| \rightarrow \infty$, the generating function 
for the conserved quantities may be written as \cite{agu}:
\bea
I_j= \int_{-\infty}^{\infty}dx \left[U_{jj}+\sum_{i\ne j}U_{ji}\Gamma_{ij}\right].
\label{ce}
\eea
It should be noted that the above expression for the conserved quantity can also be extended to the case of periodic boundary conditions. For example, within an interval of length $2a$, if the field $\psi_i (\phi_i)$ assumes the same value at the boundaries at $x=a$ and $x=-a$, Eq. (\ref{ce}) can be obtained from Eq. (\ref{CE}).
This is due to the fact that the integral of the right hand side in Eq. (\ref{CE}) vanishes even if the fields $\{\psi_i,\phi_j\}, i,j\in \{1,2\}$, satisfy a periodic boundary condition and the 
limit of the integration is taken within this period.

Since the components $U_{ij}$ and $V_{ij}$ of the matrices $U$ and $V$ are respectively given by Eqs. (\ref{u}) and (\ref{v}),
the Riccati Eqs. (\ref{rcx}) and (\ref{rct}) for the auxiliary function $\Gamma_{21}$, take the following form
\bea
\partial_x\Gamma_{21}&=&i(\lambda \psi_2-\frac{1}{\lambda}\psi_1)- \{ 2i\epsilon \rho_--\frac{i}{2}(\lambda^2-\frac{1}{\lambda^2})\} \Gamma_{21}-i\epsilon(\lambda \phi^*_2-\frac{1}{\lambda}\phi^*_1){{\Gamma}^2_{21}},
\ \ \label{21x}\\
\partial_t\Gamma_{21}&=&i(\lambda \psi_2+\frac{1}{\lambda}\psi_1)- \{ 2i\epsilon \rho_+-\frac{i}{2}(\lambda^2+\frac{1}{\lambda^2})\} \Gamma_{21}-i\epsilon(\lambda \phi^*_2+\frac{1}{\lambda}\phi^*_1){{\Gamma}^2_{21}}.
\ \ \label{21t}
\eea
In order to solve Eq. (\ref{21x}) for $\Gamma_{21}$, we expand $\Gamma_{21}$ in term of inverse power of $\lambda$,
\bea
\Gamma_{21}(x,t; \lambda)= \sum_{k=1}^{\infty}\frac{\Gamma^{(k)}(x,t)}{\lambda^{k}}.
\label{gex}
\eea
Using Eq. (\ref{21x}), the expansion coefficients $\Gamma^{(k)}$ can easily be obtained in a recursive way and have the following expression
\bea
\Gamma^{(n)}&=& 2\big(\psi_1\delta_{n3}-\psi_2\delta_{n1}\big)-2i \Gamma_{x}^{(n-2)}+4\epsilon \rho_{-}\Gamma^{(n-2)}
+\Gamma^{(n-4)}\nn\\
&+&2\epsilon \big[\phi_2^*\sum_{k=1}^{n-2}\Gamma^{(n-1-k)}\Gamma^{(k)}-\phi_1^*\sum_{k=1}^{n-4}\Gamma^{(n-3-k)}\Gamma^{(k)}\big],
\label{reG21}
\eea
with $\Gamma^{(k)}=0,\ \ \forall \ \ k<1$.
The first few expansion coefficients may be written as
\bea
\Gamma^{(1)}=-2\psi_2, \ \ \Gamma^{(2)}=0,\ \  \Gamma^{(3)}=2\psi_1+4i\psi_{2x}+8\epsilon \psi_2\psi_1 \phi^*_1.
\eea
Since $\Gamma^{(2)}=0$ and since for even $n$, each term in the right hand side of Eq. (\ref{reG21}) involves at least one even expansion coefficient, it follows interestingly that all the even expansion coefficients are zero, i.e, $\Gamma^{(2n)}=0$. As mentioned above, in order to obtain the solution for $\Gamma_{21}$, we consider only Eq. (\ref{21x}). The time derivative part, i.e., Eq. (\ref{21t}) is automatically taken into account by the compatibility condition $v_{ xt} = v_{tx}$ or in other words by the equations of motions. The generating functions for the conserved quantities, i.e., Eq.(\ref{ce}), takes the following form with $j=1$
\bea
I_1=\int_{-\infty}^{\infty}dx \left[U_{11}+U_{12}\Gamma_{21}\right].
\label{ce1}
\eea
Substituting the expression of $\Gamma_{21}$ from Eq. (\ref{gex}) into the above equation and using the explicit 
form of $U_{11}, U_{12}$ from Eq. (\ref{u}), we obtain an infinite number of conserved quantities given by the 
expression $I_1= \sum_{k=0}^{\infty}\frac{I_1^{(k)}}{\lambda^{k}}$ where each of the expansion coefficients $I_1^{(k)}$ represents a conserved quantity. 
The general expression for the conserved quantities may, therefore, be written as
\bea
I_1^{(n)}=i\epsilon\int_{-\infty}^{\infty}\big[ \rho_{-}\delta_{n0}+ \big(\phi_2^*\Gamma^{(n+1)}-\phi^*_1\Gamma^{(n-1)}\big)\big]dx.
\label{I1}
\eea
By inserting (\ref{reG21}) in the above equation, first few conserved quantities can now be written in the following form
\bea
&&I_1^{(0)}= -i\epsilon\int_{-\infty}^{+\infty} (\psi_1\phi^*_1 + \psi_2 \phi^*_2) dx, \nn \\
&&I_1^{(2)}= i\epsilon\int_{-\infty}^{+\infty}\big[4i\phi^*_2{\psi_2}_x  + 2( \phi^*_2\psi_1 + \phi^*_1 \psi_2)
+ 8\epsilon \psi_1\phi^*_1\psi_2\phi^*_2\big]dx.
\label{cc21c}
\eea
Since $\Gamma^{(2n)}=0$, it is evident from (\ref{I1}) that all odd expansion coefficients like $I_1^{2n+1}$ are zero.

Another set of conserved quantities can be obtained by expanding $\Gamma_{21}$ in term of positive powers of 
$\lambda$,
\bea
\Gamma_{21}(x,t; \lambda)= \sum_{k=1}^{\infty}\bar{\Gamma}^{(k)}(x,t)\lambda^{k}.
\label{gex1}
\eea
Using similar steps as earlier, the general expression for the expansion coefficients may be obtained as,
\bea
\bar{\Gamma}^{(n)}&=& 2\big(\psi_2\delta_{n3}-\psi_1\delta_{n1}\big)+2i \bar{\Gamma}_{x}^{(n-2)}-4\epsilon \rho_{-}\bar{\Gamma}^{(n-2)}
+\bar{\Gamma}^{(n-4)}\nn\\
&-&2\epsilon \big[\phi_2^*\sum_{k=1}^{n-4}\bar{\Gamma}^{(n-3-k)}\bar{\Gamma}^{(k)}-\phi_1^*\sum_{k=1}^{n-2}\bar{\Gamma}^{(n-1-k)}\bar{\Gamma}^{k}\big].
\label{reGb21}
\eea
where $\bar{\Gamma}^{(k)}=0$ for $k<1$. The first few expansion coefficients may be written as
\bea
\bar{\Gamma}^{(1)}=-2\psi_1, \ \ \bar{\Gamma}^{(2)}=0,\ \  \bar{\Gamma}^{(3)}=2\psi_2-4i\psi_{1x}+8\epsilon \psi_2\psi_1 \phi^*_2.
\label{cG}
\eea
Due to the same reason as mentioned earlier,  the even expansion coefficients, are zero, i.e, $\bar{\Gamma}^{(2n)}=0$.
Substituting the expression of $\Gamma_{21}$ from Eq. (\ref{gex1}) into Eq. (\ref{ce1}) and using the explicit form of $U_{11}, U_{12}$ from Eq. (\ref{u}), an infinite number of conserved 
quantities, given by the expression $I_1= \sum_{k=0}^{\infty}\bar{I}_1^{(k)}{\lambda^{k}}$, can be obtained. The general expression for the conserved quantities may be expressed as
\bea
\bar{I}_1^{(n)}=i\epsilon \int_{-\infty}^{\infty}\big[\rho_{-}\delta_{n0}+\big(\phi_2^*\bar{\Gamma}^{(n-1)}-\phi^*_1\Gamma^{(n+1)}\big)\big]dx.
\label{bI1}
\eea
By inserting (\ref{cG}) in the above equation, first few conserved quantities can now be written as
\bea
&& \bar{I}_1^{(0)}=  i\epsilon \int_{-\infty}^{+\infty} (\psi_1\phi^*_1 + \psi_2 \phi^*_2)dx \, , \nn \\
&& \bar{I}_1^{(2)}= i\epsilon \int_{-\infty}^{+\infty}\big[4i \phi^*_1 \psi_{1x}  - 2(\phi^*_1\psi_2+ \phi^*_2 \psi_1) 
- 8\epsilon \psi_2 \phi^*_2\psi_1\phi^*_1\big]dx.
\label{Cqb21}
\eea
Since $\bar{\Gamma}^{(2n)}=0$, it is evident from (\ref{bI1}) that all odd expansion coefficients like $\bar{I}_1^{2n+1}$ are zero.

By using a suitable combination of $I_1^{(n)}$ and $\bar{I}_1^{(n)}$ given in Eqs.  (\ref{cc21c}) and (\ref{Cqb21}), the following real conserved quantities are constructed,
\bea
N&=&-\frac{1}{2i\epsilon} (I_1^{(0)}-\bar{I}_1^{(0)})+C.C.= \int_{-\infty}^{+\infty} (\psi_2\phi^*_2 + \psi_1 \phi^*_1+C.C.)dx,\\
P &=&\frac{1}{4i\epsilon}(I_1^{(2)}+\bar{I}_1^{(2)})+C.C.
= \int_{-\infty}^{+\infty}\big[ i(\psi_{2x} \phi^*_2+ \psi_{1x} \phi^*_1) +C.C.\big]\, dx, \nn \\ 
H&=&\frac{1}{4i\epsilon}(I_1^{(2)}-\bar{I}_1^{(2)})+C.C.\nn \\
&=& \int_{-\infty}^{\infty}\big[i(\psi_{2x}\phi^*_2 - \psi_{1x} \phi^*_1) + (\psi_1\phi^*_2 + \psi_2 \phi^*_1) +4 \epsilon \psi_1 \phi^*_1 \psi_2  \phi^*_2+C.C.\big]dx,
\label{cqf}
\eea
where $C.C.$ denotes the complex conjugation of the preceding terms.The physical meaning and significance of these conserved quantities will be presented in the next section.

The conserved quantities for the various nonlocal models related to BMTM can easily be obtained  by using the reduction relations (\ref{r1}), (\ref{r2}), (\ref{r3}), (\ref{r4}) and (\ref{r5}),
in the expressions of $I_1^{(k)}$ and $\bar{I}_1^{(k)}$. Thus, for each of the nonlocal models, we get two sets of infinite number of conserved 
quantities. By employing a suitable combination of  $I_1^{(k)}$ and $\bar{I}_1^{(k)}$, we construct the following real conserved quantities 
for each of the nonlocal models,
\bea
&&N_i= -\frac{1}{2i\epsilon} (I_1^{(0)}-\bar{I}_1^{(0)})+C.C., \ \ 
P_i =\frac{1}{4i\epsilon}(I_1^{(2)}+\bar{I}_1^{(2)})+C.C.,\nn\\
&&H_i = \frac{1}{8i\epsilon}(I_1^{(2)}-\bar{I}_1^{(2)})+C.C.,
\eea
where $i=I, II,.., V$, corresponds to the type of the nonlocal models and for each of the choices, the corresponding 
reduction relations (\ref{r1}), (\ref{r2}), (\ref{r3}), (\ref{r4}) and (\ref{r5}) should be used in the expressions of $I_1^{(k)}$ and $\bar{I}_1^{(k)}$.

The explicit expressions of the conserved quantities $N_i$, $P_i$ and $H_i$ for the nonlocal models are presented below.\\

Type-I: Real reverse space:

\bea
N_I&=& \int_{-\infty}^{+\infty} \big[\psi_2\psi^{*}_1(-x,t) + \psi_1 \psi_2^{*}(-x,t)+C.C.\big]dx\nn \\
P_I&=& i\int_{-\infty}^{+\infty} \big[\psi_{2x} \psi^{*}_1(-x,t)+\psi_{1x}\psi^{*}_2(-x,t)+C.C.\big] dx \nn \\ 
H_I&=&  \frac{1}{2}\int_{-\infty}^{\infty}\big[i\{\psi_{2x} \psi^{*}_1(-x,t) - \psi_{1x} \psi^{*}_2(-x,t)\}\nn\\
&+&\{\psi_1\psi^{*}_1(-x,t) + \psi_2 \psi^{*}_2(-x,t)\}+4 \epsilon \psi_1 \psi_2 \psi^{*}_1(-x,t) \psi^{*}_2(-x,t)+C.C.\big]
dx. \nn\\
\label{1cqf}
\eea

Type-II: Real reverse time:

\bea
N_{II}&=& \int_{-\infty}^{+\infty} \big[\psi_2\psi^{*}_1(x,-t) - \psi_1 \psi_2^{*}(x,-t)+C.C.\big]dx\nn \\
P_{II}&=& i\int_{-\infty}^{+\infty} \big[\psi_{2x} \psi^{*}_1(x,-t)- \psi_{1x}\psi^{*}_2(x,-t)+C.C.\big]\, dx \nn \\ 
H_{II}&=&  \frac{1}{2}\int_{-\infty}^{\infty}\big[i\{\psi_{2x} \psi^{*}_1(x,-t) + \psi_{1x} \psi^{*}_2(x,-t)\}\nn \\
&+&\{\psi_1\psi^{*}_1(x,-t) - \psi_2 \psi^{*}_2(x,-t)\}-4 \epsilon \psi_1 \psi_2 \psi^{*}_1(x,-t) \psi^{*}_2(x,-t)+C.C.\big]\,
dx. \nn\\
\label{2cqf}
\eea

Type-III: Complex reverse time:

\bea
N_{III}&=& \int_{-\infty}^{+\infty} \big[\psi_2\psi_1(x,-t) + \psi_1 \psi_2(x,-t)+C.C.\big]dx\nn \\
P_{III}&=&i\int_{-\infty}^{+\infty} \big[\psi_{2x} \psi_1(x,-t)+ \psi_{1x}\psi_2(x,-t)+C.C.\big]\, dx \nn \\ 
H_{III}&=&  \frac{1}{2}\int_{-\infty}^{\infty}\big[i\{\psi_{2x} \psi_1(x,-t) - \psi_{1x} \psi_2(x,-t)\}\nn \\
&+&\{\psi_2\psi_2(x,-t) + \psi_1 \psi_1(x,-t)\}+4 \epsilon \psi_1 \psi_2 \psi_1(x,-t) \psi_2(x,-t)+C.C.\big]\,
dx. \nn\\
\label{3cqf}
\eea

Type-IV: Complex reverse space:

\bea
N_{IV}&=& \int_{-\infty}^{+\infty} \big[\psi_2\psi_1(-x,t) - \psi_1 \psi_2(-x,t)+C.C.\big]dx\nn \\
P_{IV}&=& i\int_{-\infty}^{+\infty} \big[\psi_{2x} \psi_1(-x,t)- \psi_{1x}\psi_2(-x,t)+C.C.\big]\, dx \nn \\ 
H_{IV}&=&  \frac{1}{2}\int_{-\infty}^{\infty}\big[i\{\psi_{2x} \psi_1(-x,t) + \psi_{1x} \psi_2(-x,t)\}\nn \\
&+&\{\psi_1\psi_1(-x,t) - \psi_2 \psi_2(-x,t)\}-4 \epsilon \psi_1 \psi_2 \psi_1(-x,t) \psi_2(-x,t)+C.C.\big]\,
dx. \nn\\
\label{3cqf1}
\eea

Type-V: Complex reverse space time:

\bea
N_{V}&=& \int_{-\infty}^{+\infty} \big[\psi_1\psi_1(-x,-t) + \psi_2 \psi_2(-x,-t)+C.C.\big]dx\nn \\
P_{V}&=& i\int_{-\infty}^{+\infty} \big[(\psi_{1x} \psi_1(-x,-t)+ \psi_{2x}\psi_2(-x,-t) )+C.C.\big]\, dx \nn \\ 
H_{V}&=& \frac{1}{2} \int_{-\infty}^{\infty}\big[i\{\psi_{2x} \psi_2(-x,-t) - \psi_{1x} \psi_1(-x,-t)\}\nn \\
&+&\{\psi_1\psi_2(-x,-t) + \psi_2 \psi_1(-x,-t)\}+4 \epsilon \psi_1 \psi_2 \psi_1(-x,-t) \psi_2(-x,-t)+C.C.\big]\,
dx. \nn\\
\label{3cqf2}
\eea
Thus, for each of the nonlocal models related to coupled BMTM, we obtain an infinite number of conserved quantities with explicit expressions 
for some of the useful ones. The existence of such infinite number of conserved quantities confirms the integrability of the systems. The next 
step is to study the PB relations among the various conserved quantities to infer about the complete integrability of the systems in the Liouville sense.

\section{Hamiltonian formulation of coupled BMTM and some of the related nonlocal systems}

In this section we first exploit the symmetry properties of coupled BMTM and related nonlocal models
to obtain some of the corresponding conserved quantities by using Noether's theorem. Next, we consider
the Hamiltonian formulation for the coupled BMTM and two of the related models having space inverted nonlocality. 
After the scale transformation (\ref{cv}),
the Lagrangian density for the coupled BMTM as given in Eq. (\ref{L}) can be written in the component form as,
\bea
{\cal L}&=&\sum^2_{j=1}\frac{i}{2}\big[\big(\phi^*_j\psi_{jt}-\psi_j\phi^*_{jt}\big)-(-1)^j\big(\phi^*_j\psi_{jx}-\psi_j\phi^*_{jx}\big)\big]\nn \\
&&-\big(\psi_2\phi^*_1 + \psi_1 \phi^*_2\big)-4 \epsilon \psi_1 \psi_2 \phi^*_1 \phi^*_2 + C.C.
\label{CBT}
\eea
It should be noted that the action for the coupled BMTM, i.e., ${\cal A}= \int {\cal L} dx dt$ is invariant under space-time translation and 
global $U(1)$ gauge transformation. Therefore, the corresponding conserved quantities can be obtained from the symmetry principle by calculating the Noether 
charge associated with these symmetries. In particular, the energy momentum tensor has the expression
\bea
T^{\mu \nu}= \sum_{j=1}^2\big[\frac{\partial {\cal L}}{\partial ({\partial_{\mu}\psi_j})}\partial^{\nu}\psi_j
+\frac{\partial {\cal L}}{\partial ({\partial_{\mu}\phi_j})}\partial^{\nu}\phi_j+ C.C. 
\big]
-g^{\mu\nu}{\cal L}\nn
\eea
where $x^{\mu}=(t,x), \partial_{\mu}=(\frac{\partial}{\partial t}, \frac{\partial}{\partial t}), g^{\mu\nu}=dia(1,-1); \mu,\nu=0,1$. 
The Hamiltonian of the coupled BMTM can be obtained from the Noether charges associated with the time translation 
symmetry and is given by the expression $H=\int T^{00}dx$. In the explicit form the Hamiltonian may be written as
\bea
 H&=& \int\big[ \big(\sum_{j=1}^2\psi_{jt}\Pi_{\psi_j}+\phi_{jt}\Pi_{\phi_j}+\psi^*_{jt}\Pi_{\psi^*_j}+\phi^*_{jt}\Pi_{\phi^*_j}\big)-{\cal L}\big]dx \nn \\
&=& \int_{-\infty}^{\infty}\big[i\big(\psi_{2x}\phi^*_2 - \psi_{1x} \phi^*_1\big) +\big(\psi_2\phi^*_1 + \psi_1 \phi^*_2\big) +4 \epsilon \psi_1 \psi_2 \phi^*_1 \phi^*_2+ C.C.] 
dx.
\label{ham}
\eea
The conserved quantity $P$, referred to as the conserved linear momentum,
is associated with the invariance of the action under spatial translation and has the following form
\bea
P =\int T^{01}dx= \int_{-\infty}^{+\infty} \big[i(\psi_{1x} \phi^*_1+ \psi_{2x}\phi^*_2)+C.C.\big]
 \, dx.
\label{Ph}
\eea
The global $U(1)$ symmetry of the action is related to its invariance under the transformation,
\bea
\psi_j \rightarrow e^{-i\theta}\psi_j, \ \ \phi_j \rightarrow e^{-i\theta}\phi_j, \ \ j=1,2
\eea
and the corresponding conserved quantity, sometimes refer to as the conserved charge, has the form
\bea
N = \int_{-\infty}^{+\infty} (\,\psi_1\phi^*_1 + \psi_2 \phi^*_2 
+ C.C.\,) dx .
\label{Nh}
\eea
Thus, completely from the symmetry arguments, here we reproduce some of the conserved quantities associated with 
coupled BMTM and also understand their physical origin.

The Lagrangian for the nonlocal systems related to coupled BMTM can readily be obtained by inserting the reductions (\ref{r1}), (\ref{r2}), (\ref{r3}), (\ref{r4}) and (\ref{r5})
respectively in the expression (\ref{CBT}). The conserved charges associated with space-time translational symmetries for these
nonlocal models can be obtained similarly, the explicit expressions of which
are given by Eqs. (\ref{1cqf}), (\ref{2cqf}), (\ref{3cqf}), (\ref{3cqf1}) and (\ref{3cqf2}) respectively (up to some overall constant factor).
Likewise, for type-I and type-II systems the conserved quantities $N_{I}$ and $N_{II}$, as presented respectively in Eqs. (\ref{1cqf}) and (\ref{2cqf}),
can be obtained from their invariance under global $U(1)$ gauge transformation.

Let us now define the fundamental equal time PB relations for the coupled BMTM in the following manner,
\bea
&& \{\psi_j (x), \phi^*_k(y)\}= \{\phi_j (x), \psi^*_k(y)\}=-i\delta(x-y)\delta_{jk},\nn\\
&& \{\psi^*_j (x), \phi_k(y)\}= \{\phi^*_j (x), \psi_k(y)\}=i\delta(x-y)\delta_{jk};\ \  j,k=1,2,
\label{dpb}
\eea
where only the non zero PB relations are mentioned.
By using these PB relations, it is easy to check that Eqs. (\ref{qs}) and (\ref{rs}) can be written concisely in the following form
\bea
\psi_{jt}=\{\psi_{j},H\}, \ \ \phi_{jt}=\{\phi_{j},H\},
\eea
where the Hamiltonian $H$ is given by Eq. (\ref{ham}). This confirms that the dynamics of the coupled BMTM is Hamiltonian in nature.
In order to construct the Hamiltonian formulation for
the nonlocal systems, it should be noted that for the time reversed nonlocal models II, III and space-time reversed system V, the equal time PB relations can not be defined
between the conjugate variables and therefore we are not considering the PB relations for these systems. It turns out that only 
 the systems of type I and IV, give consistent fundamental equal time PB relations which can reproduce the equations of motion.
For example, the nonzero equal time PB relations for type-I system may be defined as,
\bea
&& \{\psi_1(x), \psi^{*}_2(-y)\}=\{\psi_2(x), \psi^{*}_1(-y)\} = -i\delta(x-y).
\label{1pb}
\eea
Therefore, in this case we can infer that the dynamics of the system is Hamiltonian with $(\psi_1(x,t), \psi^{*}_2(-x,t))$ and $(\psi_2(x,t), \psi^{*}_1(-x,t))$ 
being two sets of canonical variables and $H_{I}$ in Eq. (\ref{1cqf}) being the Hamiltonian. Indeed, by using PB relations (\ref{1pb}), the equations of motions (\ref{qse1})
can be written in the following form,
\bea
\psi_{1t}= \{\psi_1, H_{I}\},\ \ \ \  \psi_{2t}= \{\psi_2, H_{I}\}.
\eea
Finally, the nonzero fundamental equal time PB relations for type-IV system 
may be  defined as, 
\bea
&&- \{\psi_1(x,t), \psi_2(-y,t)\}= \{\psi^*_1(x,t), \psi^*_2(-y,t)\}=- i\delta(x-y).
\label{3pb}
\eea
By using the PB relations (\ref{3pb}), the equations of motions (\ref{qse4}) can now be
written in a concise manner as, 
\bea
\psi_{1t}= \{\psi_1, H_{IV}\},\ \ \ \  \psi_{2t}= \{\psi_2, H_{IV}\}. 
\eea
Thus for the Hamiltonian $H_{IV}$ given by Eq. (\ref{3cqf1}), the dynamics of the system is recovered by
taking $(\psi_1(x,t), -\psi_2(-x,t))$ and $(\psi^*_1(x,t), \psi^*_2(-x,t))$ as canonical variables.

It may be noted that the coupled BMTM as well as related space inverted nonlocal models
are constrained systems having second class constraints. 
Therefore, a systematic treatment of constructing the equal time PB relations for these models can be done by
employing Dirac’s method for constrained systems. Indeed, the equal time PB relations considered in this section can be recovered by using the 
Dirac bracket formulation.


\section{Complete integrability of coupled BMTM and type I nonlocal model }

In this section, at first we investigate the complete integrability of the coupled BMTM.
In fact, by using the fundamental PB relations for coupled BMTM, it is easy to check that the conserved quantities 
such as the Hamiltonian, charge and momentum have vanishing PB relations among themselves.
This result strongly suggests that the infinite set of conserved
quantities obtained for this model are in involution, implying
completely integrablility of the system in the Liouville sense.
In this context it may be noted that, the PB relations among various elements 
of the classical monodromy matrix for the original BMTM has been considered earlier to establish 
the complete integrability of the model in the Liouville sense  \cite{ta,kul}. 
In the present case, we first define the monodromy matrix for the coupled BMTM and show that, under vanishing boundary conditions
on the fields at $|x|\rightarrow \infty$, the diagonal elements of this monodromy matrix act as the generator of the infinite hierarchy of 
conserved quantities obtained in section 3. In the next step,
the PB relations among the  elements of monodromy matrix for
coupled BMTM having four independent complex fields are derived to get the PB relations between the conserved quantities. 

We start by recalling the first equation of (\ref{uvn}) which may be written as,
\bea
\partial_{x_2}v(x_2)=U(x_2,\lambda)v(x_2),
\eea
and define the monodromy matrix at a finite interval $T(x_2,x_1,\lambda)$ for the coupled BMTM by the following
relation
\bea
v(x_2)=T^{x_2}_{x_1}(\lambda)v(x_1).
\label{5.2}
\eea
Using Eq. (\ref{5.2}), it is easy to check that the monodromy matrix satisfies the equation
\bea
\big(\partial_{x_2}-U(x_2,\lambda)\big)T^{x_2}_{x_1}(\lambda)=0
\label{e1}
\eea
with the condition $T^{x_2}_{x_2}(\lambda)= I$, where $I$ is the $2\times 2$ unit matrix.
A formal solution of (\ref{e1}) is given by
\bea
T^{x_2}_{x_1}(\lambda)=P \exp{\int_{x_1}^{x_2}}U(x,\lambda)dx,
\label{TU1}
\eea
where $P$ denotes the path ordering. Following the procedure as outlined in Ref. \cite{YB},
the monodromy matrix at infinite interval can be defined for the coupled BMTM as,
\bea
T(\lambda)\equiv 
\bp
a(\lambda) & c(\lambda)\\
b(\lambda) & d(\lambda)
\ep
=\lim_{\stackrel{x_2 \rightarrow +\infty}{x_1
\rightarrow -\infty}}  e(-x_2,\lambda) T_{x_1}^{x_2}(\lambda) e(x_1,\lambda)
\label{T2}
\eea
 where $e(x,\lambda)= \exp{[-\frac{i}{4}(\lambda^2-\frac{1}{\lambda^2})\sigma_3x]}$. 
Now, in order to solve Eq. (\ref{e1}), it is convenient to decompose the equation in a diagonal and non diagonal part. To fulfill this purpose, the Lax operator  (\ref{u}) is expressed as $U(x, \lambda ) = U_d( x, \lambda ) + U_{nd}( x, \lambda )$, where $U_d(x, \lambda)$ is the diagonal part and $U_{nd}(x, \lambda)$ is the non-diagonal part of $U(x,\lambda)$
and it is assumed that the monodromy matrix $T_{x_1}^{x_2}(\lambda)$ can be written in the following form,
\bea
T_{x_1}^{x_2}(\lambda ) = \Big(\, 1+ W( x_2, \lambda )\, \Big)\exp Z( x_2, x_1
, \lambda )\Big(\, 1+ W( x_1, \lambda)\,\Big)^{-1} \, ,
\label{e2}
\eea
where $Z( x_2, x_1, \lambda )$ is a diagonal matrix and $W(x, \lambda )$ is a non diagonal one
satisfying the condition $W(x, \lambda) \rightarrow 0$ at $|x| \rightarrow \infty $. Eq. (\ref{e1}) can now be decomposed into:
\bea
&&\frac{dZ}{dx_2}= U_d + U_{nd}W \, 
\label{Z}\\
&&\frac{dW}{dx_2} - 2U_d W - U_{nd} + WU_{nd}W = 0,
\label{W}
\eea
where local matrices $U_{d}$, $U_{nd}$ and $W$ depends on the variable $x_2$. Further, without loss of generality, we write $W(x_2, \lambda)$ and
$Z(x_2,x_1, \lambda)$ in the form
\bea
&&W( x_2, \lambda ) = -\epsilon w_1^*(x_2, \lambda )\sigma_+ + w_2( x_2, \lambda )\sigma_- \,  \\  
&& Z( x_2, x_1, \lambda )= z_1(x_2,x_1,\lambda)\frac{1}{2}(I+\sigma_3)-z_2(x_2,x_1,\lambda)\frac{1}{2}(I-\sigma_3).
\eea
Substituting Eq. (\ref{e2}) in the expression (\ref{T2}) one can obtain
the following relations
\bea 
\ln a(\lambda) &=& \lim_{\stackrel{x_2 \rightarrow +\infty}{x_1
\rightarrow -\infty}} \big \{ z_1(x_2,x_1,\lambda) + \frac{i}{4}(\lambda^2-\frac{1}{\lambda^2})(x_2-x_1) \big \}
\label{lna}\\
\ln d(\lambda) &=& \lim_{\stackrel{x_2 \rightarrow +\infty}{x_1
\rightarrow -\infty}} \big \{ -z_2(x_2,x_1,\lambda) - \frac{i}{4}(\lambda^2-\frac{1}{\lambda^2})(x_2-x_1) \big \}.
\label{lnd}
\eea
 Solving Eq. (\ref{Z}) for $z_1$ and $z_2$ and substituting the explicit form of $z_1$ and $z_2$ 
into the above expressions, we get the following form of $\ln a(\lambda)$ and $\ln d(\lambda)$ respectively,
\bea
\ln a(\lambda ) &=& i\epsilon\int_{-\infty}^{+\infty} \rho_- dx + i\epsilon\lambda\int_{-\infty}^{+\infty}\phi^*_2 w_2 dx -
\frac{i\epsilon}{\lambda} \int_{-\infty}^{+\infty}\phi^*_1 w_2 dx,
\label{lna1}\\
\ln d(\lambda ) &=& -i\epsilon\int_{-\infty}^{+\infty} \rho_- dx - i\epsilon\lambda\int_{-\infty}^{+\infty}\psi_2 w^*_1 dx +
\frac{i\epsilon}{\lambda} \int_{-\infty}^{+\infty}\psi_1 w_1^* dx.
\label{lnd2}
\eea
By using Eq. (\ref{W}), it is easy to check that $w_2$ satisfies a Riccati equation of the form
\bea
\frac{\partial w_2}{\partial x}=i \big(\lambda \psi_2-\frac{1}{\lambda}\psi_1\big)-\big[2i\epsilon \rho_{-}-
\frac{i}{2}\big(\lambda^2-\frac{1}{\lambda^2}\big)\big] w_2-i \epsilon \big(\lambda \phi^*_2-\frac{1}{\lambda}\phi_1^*\big) w_2^2.
\label{exw_2}
\eea
Next, we expand $w_2( x,\lambda )$ in inverse powers of $\lambda$ as
\bea
w_2( x, \lambda ) = \sum_{j=0}^\infty
\frac{v_j}{\lambda^{2j+1}},
\label{w_2}
\eea
and by employing Eq.(\ref{exw_2}) find that the expansion coefficients $v_j$ satisfy the following recursion relation
\bea
v_{n}&=& 2\big(\psi_1\delta_{n1}-\psi_2\delta_{n0}\big)-2i v_{(n-1)x}+4\epsilon \rho_{-}v_{n-1}
+v_{n-2}\nn\\
&+&2\epsilon \big[\phi_2^*\sum_{k=0}^{n-1}v_{n-1-k}v_{k}-\phi_1^*\sum_{k=0}^{n-3}v_{n-2-k}v_{k}\big],
\label{reG21v}
\eea
where $n=0,1, 2...$. and $v_n=0$ for $n<0$. By using this recursion relations, the first few expansion coefficients may be explicitly 
written as
\bea
v_0 = -2\psi_2,\ \  v_1 = 2\psi_1+4i\psi_{2x} +
8\epsilon (\psi_1\phi^*_1) \psi_2.\nn
\eea
Substituting Eq. (\ref{w_2}) in the expression of $\ln a(\lambda)$ as given by Eq. (\ref{lna1}), one gets 
the following relation
\bea
\ln a( \lambda ) = \sum_{n=0}^{\infty} \frac{C_n}{\lambda^{2n}},
\label{lnaC}
\eea
where the expansion coefficients $C_n$'s have the following expression,
\bea
C_{n}=i\epsilon \int_{-\infty}^{\infty}\big[\rho_{-}\delta_{n0}+\big(\phi_2^*v_{n}-\phi^*_1v_{n-1}\big)\big]dx.
\label{bI1C}
\eea
The first two of them may be explicitly written as
\bea
&&C_0 = -i\epsilon\int_{-\infty}^{+\infty}\big(\,\psi_1\phi^*_1 + \psi_2 \phi^*_2 
\,\big) dx  \nn \\
&& C_1 =i\epsilon\int_{-\infty}^{+\infty} \big[4i\phi^*_2\psi_{2x}  + 2
\big( \phi^*_2\psi_1 +\phi^*_1 \psi_2\big) 
+ 8\epsilon \psi_1\phi^*_1\psi_2\phi^*_2\big]dx.
\nn
\eea
It is evident that the coefficients $C_n$ coincide with the conserved quantities $I^{(2n)}_1$ 
obtained in section 3. 

\noindent Next we expand $w_2( x, \lambda )$ in positive powers of $\lambda$ as,
\bea
w_2(x, \lambda ) = \sum_{j=0}
^\infty \tilde{v}_j \lambda^{2j+1}.
\label{w_2p}
\eea
In a similar way as outlined above, it can be shown that the expansion coefficients $\tilde{v}_j$ satisfy the recursion relations
\bea
\tilde{v}_{n}&=& 2\big(\psi_2\delta_{n1}-\psi_1\delta_{n0}\big)+2i \tilde{v}_{(n-1)x}-4\epsilon \rho_{-}\tilde{v}_{n-1}
+\tilde{v}_{n-2}\nn\\
&+&2\epsilon \big[\phi_1^*\sum_{k=0}^{n-1}\tilde{v}_{n-1-k}\tilde{v}_{k}-\phi_2^*\sum_{k=0}^{n-3}\tilde{v}_{n-2-k}\tilde{v}_{k}\big],
\label{reG21tv}
\eea
where $n=0,1,2...$ and $\tilde{v}_j=0$ for $n<0$. By using (\ref{reG21tv}),
the first few $\tilde{v}_j$'s can be written as 
\bea
\tilde{v}_0=-2\psi_1 \,\,, \tilde{v}_1 =2\psi_2  -4i\psi_{1x} 
+ 8\epsilon(\phi^*_2\psi_2)\psi_1.
\eea
Substituting Eq. (\ref{w_2p}) in the expression of $\ln a(\lambda)$ as given by Eq. (\ref{lna1}), one gets the series
\bea
\ln a(\lambda)= \sum_{n=0}
^\infty \tilde{C}_n\lambda^{2n},
\label{lnatC}
\eea
where $\tilde{C}_n$s are the expansion coefficients having the following form
\bea
 \tilde{C}_{n}=i\epsilon \int_{-\infty}^{\infty}\big[\rho_{-}\delta_{n0}-\big(\phi_1^*\tilde{v}_{n}-\phi^*_2\tilde{v}_{n-1}\big)\big]dx.
\label{bI1C}
\eea
The explicit expressions of the first few expansion coefficients are as follows,
\bea
&& \tilde{C}_0 =  i\epsilon\int_{-\infty}^{+\infty} \big(\,\psi_1\phi^*_1 + \psi_2 \phi^*_2 \,\big) dx , \nn \\
&& \tilde{C}_1 =i\epsilon \int_{-\infty}^{+\infty}\big[4i\phi^*_1\psi_{1x} - 2
\big( \phi^*_2\psi_1 + \phi^*_1 \psi_2 \big)
- 8\epsilon \psi_1 \phi^*_1\psi_2\phi^*_2\big]dx.
\eea
Again, the expansion coefficients $\tilde{C}_n$ coincide with the conserved quantities $\bar{I}^{(2n)}_1$ obtained in section 3.
Hence, by using $C_n$ and $\tilde{C}_n$ in place of ${I}^{(2n)}_1$ and $\bar{I}^{(2n)}_1$ respectively in Eq. (\ref{cqf}),
it is possible to construct  the conserved quantities $N, P$ and $H$
which are respectively associated with the charge, momentum and Hamiltonian of the coupled BMTM.

Since the diagonal element $a(\lambda)$ of the monodromy matrix 
generate the infinite hierarchy of conserved quantities obtained in section 3,
the obvious next step for establishing the completely integrability of the coupled BMTM
is to show that $\{a(\lambda), a(\mu)\}=0$ for all possible values of $\lambda$ and $\mu$.
To this end, we consider  the following ultralocal PB relation
among the elements of the Lax operator for the systems under investigation \cite{fd, kul}
\bea
\left\{ U( x, \lambda ) {\stackrel {\otimes}{,}}
 U( y, \mu ) \right\} = \left [ r( \lambda, \mu ), U( x,
\lambda ) \otimes  I + I \otimes U( y, \mu ) \right ]\, \delta( x-y ) \, ,
\label{U1}
\eea 
where $U$ is the matrix associated with the space derivative part of the
Lax equations and $r(\lambda, \mu)$ is a matrix of appropriate dimension with $c$-number valued elements. 
Since in the present case $U$ is given by Eq. (\ref{u}), using the PB relation
(\ref{dpb}) we find that Eq. (\ref{U1}) is satisfied for $r( \lambda, \mu )$ given by
\bea
r( \lambda, \mu ) = -\epsilon \left \{ \, t^c\sigma_3\otimes\sigma_3 + s^c (
\sigma_+\otimes\sigma_- + \sigma_-\otimes\sigma_+ ) \, \right\}
\label{1r1}
\eea
with $ t^c = \frac{\lambda^2 + \mu^2}{2( \lambda^2 - \mu^2 )} ,\quad s^c =
\frac{2\lambda\mu}{\lambda^2 - \mu^2} $, and $\sigma_i, i=1,2,3$, are the three Pauli matrices where
$\sigma_{\pm}=\frac{1}{2}\big(\sigma_1\pm\sigma_2\big)$. It may be noted that, Eq. (\ref{U1}) implies the following commutation relation
for the monodromy matrices in the finite interval \cite{fd},
\bea
\left\{ T_{x_1}^{x_2}( \lambda ) {\stackrel {\otimes}{,}}
 T_{x_1}^{x_2}(\mu ) \right\} = \left [ r( \lambda, \mu ), T_{x_1}^{x_2}( \lambda ) \otimes T_{x_1}^{x_2}(\mu )  \right ].
\label{T0}
\eea 
which is sometimes referred to as the classical Yang-Baxter equation.
Hence, by using Eq. (\ref{T2}) and employing the expression of $r( \lambda, \mu )$ given in Eq. (\ref{1r1}),
we obtain the classical Yang-Baxter equation for the monodromy matrix in the infinite 
intervals as
\bea
\left\{ T( \lambda ){\stackrel{\otimes}{,}} T( \mu ) \right\} 
= r_+ ( \lambda, \mu ) T( \lambda )
\otimes T( \mu ) - T( \lambda ) \otimes T ( \mu )r_-( \lambda, \mu ) \, ,
\label{T1}
\eea
where 
\bea
r_\pm = -\epsilon \left ( t^c \sigma_3 \otimes \sigma_3 + s^c_\pm \sigma_+ 
\otimes \sigma_- + s^c_\mp \sigma_- \otimes \sigma_+ \right ) \, ,
\nn
\label{2r2}
\eea 
with $ s^c_\pm = \pm 2 i \pi \lambda^2 \delta ( \lambda^2 - \mu^2 )$.  By applying Eq. (\ref{T2}) and Eq. (\ref{T1}), 
it is now easy to obtain the PB relations among the elements of the monodromy matrix as,
\bea
&&\{ a( \lambda ) , a( \mu ) \} = 
 \{ b( \lambda ) , b( \mu ) \} = \{ c( \lambda ) , c( \mu ) \}=\{ d( \lambda ) , d( \mu ) \} =\{ a( \lambda ) , d( \mu ) \}=0,\nn \\
&&\{ a( \lambda ) , b( \mu ) \} = \epsilon \left(
\frac{\lambda^2 + \mu^2 }{\lambda^2 - \mu^2} \right) \,
a( \lambda )b( \mu ) - 2i\pi \epsilon \lambda^2 \, \delta( \lambda^2 - \mu^2 ) \, 
b(\lambda )a(\mu)  \, ,\nn \\
&&\{ a( \lambda ) , c( \mu ) \} = -\epsilon \left(
\frac{\lambda^2 + \mu^2 }{\lambda^2 - \mu^2} \right) \,
a( \lambda )c( \mu ) + 2i\pi \epsilon \lambda^2 \, \delta( \lambda^2 - \mu^2) \,
c(\lambda )a( \mu ) \, , \nn \\
&&\{ b( \lambda ) , c( \mu ) \} 
= 4 \pi i \epsilon \lambda^2 \, \delta( \lambda^2 -
\mu^2) \,  {a( \lambda )}{d( \lambda )} \nn \\
&& \{ b( \lambda ) , d( \mu ) \}=  -\epsilon \left(
\frac{\lambda^2 + \mu^2 }{\lambda^2 - \mu^2} \right) \,
b( \lambda )d( \mu ) +2i\pi \epsilon \lambda^2 \, \delta( \lambda^2 - \mu^2 ) \, 
d(\lambda )b(\mu)   \nn\\
&& \{ c( \lambda ) , d( \mu ) \}= \epsilon \left(
\frac{\lambda^2 + \mu^2 }{\lambda^2 - \mu^2} \right) \,
c( \lambda )d( \mu ) -2i\pi \epsilon \lambda^2 \, \delta( \lambda^2 - \mu^2 ) \, 
d(\lambda )c(\mu).
\label{pbt}
\eea
Due to the relation $\{ a( \lambda ) , a( \mu ) \} = 0$, it immediately follows that all the expansion coefficients  appearing in the expressions (\ref{lnaC}) and (\ref{lnatC}) 
of $\ln{a}(\lambda)$, will Poisson commute with each other, i.e., 
\bea
\{C_m, C_n\}=\{\tilde{C}_m, \tilde{C}_n\}=\{C_m, \tilde{C}_n\}=0
\label{cncm}
\eea
for all possible values of $m$ and $n$.  
 For the coupled BMTM, the complex conjugates
$C_n^*$ and $\tilde{C}_n^*$ form another  two sets of conserved quantities. Eq. (\ref{cncm}) together with Eq. (\ref{dpb}) implies that
\bea
\{C^*_m, C^*_n\}=\{\tilde{C}^*_m, \tilde{C}^*_n\}=\{C^*_m, \tilde{C}^*_n\}=0.
\label{ccncm}
\eea
Since ${C_n}$, $\tilde{C}_n$ are functional of
the fields $(\psi_1,\psi_2, \phi_1^*,\phi_2^*)$ and ${C^*_n}$, $\tilde{C}^*_n$ are functional of
the fields  $(\psi^*_1,\psi^*_2, \phi_1,\phi_2)$, by using the PB relations  (\ref{crpb}) it is easy to check that
\bea
\{C_m, C^*_n\}=\{\tilde{C}_m, \tilde{C}^*_n\}=\{C_m, \tilde{C}^*_n\}=0.
\label{crpb}
\eea
Thus  for the coupled BMTM it follows that all the conserved quantities Poisson commute with each other, which confirms the integrability of the system in the Liouville sense. 

We now apply the procedure outlined above in the case of type I nonlocal system to study its complete integrability.
 By substituting the reduction (\ref{r1}) in Eq. (\ref{u}),
 the $U$ matrix in this case is obtained as,
\bea
U&=&
\bp
i\epsilon \rho_--\frac{i}{4}(\lambda^2-\frac{1}{\lambda^2})\ \  & i\epsilon\big(\lambda \psi^*_1(-x,t)-\frac{\psi^*_2(-x,t)}{\lambda}\big)\\
i(\lambda \psi_2-\frac{\psi_1}{\lambda}) & -i\epsilon\rho_{-}+\frac{i}{4}(\lambda^2-\frac{1}{\lambda^2})
\ep,
\label{utype1}
\eea
where $\rho_{-}=\psi_1^*(-x,t)\psi_2-\psi_2^*(-x,t)\psi_1$. By using the fundamental PB relations (\ref{1pb}), 
 it is easy to check that this $U$ matrix satisfies the ultralocal PB relation (\ref{U1}) with the same $r$ matrix (\ref{1r1}) as in 
the case of coupled BMTM. Hence, it is evident that for this nonlocal model the elements of the monodromy matrix satisfy 
the same PB relations as given by Eq. (\ref{pbt}). Therefore,
exactly like the case of coupled BMTM, all the conserved quantities for this nonlocal model Poisson commute among themselves, 
which confirms the complete integrability of the system. Finally we like to comment that, for the $U$ matrix corresponding to the
nonlocal system of type IV, we could not find any $r$ matrix which satisfies the ultralocal PB relation (\ref{U1}). Hence, it seems 
that type IV system with the PB relations (\ref{3pb}) may not be completely integrable.

\section{Summary and discussions}

A coupled BMTM involving two interacting Dirac spinors has been introduced in the present paper. 
This coupled BMTM is Lorentz invariant and respects many other important symmetries, such as parity, time reversal and global $U(1)$ gauge 
transformation of the original BMTM. The Lagrangian and Hamiltonian formulation for this system has been presented. 
Further, in order to study the integrable properties of this system, the Lax pair is constructed and the equations of motion
are obtained from the zero curvature condition. By using the linear equations associated with the Lax pair, it has been shown
that the system possesses an infinite number of conserved 
quantities which confirm the integrability of the system. A detail investigation has also been done on the complete integrability 
of the coupled BMTM. In particular, by using the ultralocal PB relations among the elements of the Lax operator,
it has been shown that the corresponding monodromy matrix satisfies the classical Yang-Baxter equation. From  this construction it follows
that the conserved quantities of coupled BMTM are in involution and hence the system is completely integrable in the Liouville sense.

An important outcome of the present investigation is the emergence  of five novel nonlocal integrable systems related to BMTM. More precisely, it has been shown that various consistent reductions between the field components of the coupled BMTM can be used to generate the above mentioned  nonlocal integrable models.
It turns out that three of the nonlocal models remain invariant under discrete transformations such as parity and time reversal. 
However, all these  nonlocal models respect combined parity-time reversal symmetry.
Further, one of these models remains invariant under proper Lorentz transformation and two other models respect global $U(1)$ gauge symmetry. 
An infinite number of conserved quantities have been obtained for each of the nonlocal models confirming the integrability of the systems. 
The Hamiltonian formulation for two of the 
nonlocal models, namely for type-I and IV systems, has been presented. 
For the system of type I, it turns out that the Yang-Baxter equation and 
the PB relations among the elements of the monodromy matrix have the same form as those of in the case of coupled BMTM, which implies
the completely integrability of this nonlocal system in the Liouville sense.

The results obtained in this paper may play a key role in stimulating some
future direction of studies which are outlined as follows. Since integrable systems often lead to
exact soliton like solutions which have important physical applications, an obvious follow up of the present investigation
would be to obtain such exact solution of the coupled BMTM and related nonlocal models. At present, work is being done in this direction by
employing some well known procedure of solving integrable nonlinear dynamical systems such as 
 inverse scattering method \cite{bmds1} and Hirota's bilinear method.  
 Further, as mentioned earlier in the introduction, there exits two versions of the
integrable classical  MTM, namely, the BMTM in which case the fields take value in the complex number field and the grassmannian
MTM (GMTM) which is obtained by considering the fields as functions taking value in the Grassmann algebra.
 In the present paper, the classical integrability of the bosonic version of the coupled MTM and the related nonlocal models are
considered. The study of these models while the fields take value in the Grassmann algebra is being pursued in Ref. \cite{gmtm}.
Finally, both the bosonic and grassmannian versions of coupled MTM in presence of defect or balanced loss and gain
might turn out to be interesting models for future study.

\section{Acknowledgements}

The authors would like to thank A. Gonz\'alez-L\'opez and F. Finkel for valuable discussions.
DS acknowledges a research fellowship from CSIR (ACK No: 362103/2k19/1, File No 09/489(0125)/2020-EMR-I).

\end{document}